\documentclass[prl,twocolumn,showpacs,preprintnumbers,amsmath,amssymb,groupedaddress,superscriptaddress,floatfix]{revtex4}
\usepackage{graphicx}
\begin{document}
\title{Muons as Local Probes of Three-body Correlations in the Mixed State
of Type-II Superconductors}

\author{G. I. Menon}
\affiliation{The Institute of Mathematical Sciences, C.I.T. Campus,  Taramani,
Chennai 600 113, India}

\author{A. Drew}\altaffiliation[Present address: ]{D\'epartement de Physique, Universit\'e de Fribourg, Chemin Du Mus\'ee 3, Fribourg, CH1700, Switzerland.}
\author{U.K. Divakar}
\author{S.L. Lee}
\affiliation{School of Physics and Astronomy, University of St. Andrews,
Fife KY16 9SS, UK}

\author{R. Gilardi}
\author{J. Mesot}
\affiliation{Laboratory for Neutron Scattering, ETH Zurich and PSI Villigen,
CH-5232 Villigen PSI, Switzerland}

\author{F.Y. Ogrin}
\affiliation{Department of Physics, University of Exeter, Exeter, EX4 4QL,
UK}

\author{D. Charalambous}\altaffiliation[Present address: ]{Meteorological Service, Ministry of Agriculture, 28 Nikis Avenue, 1086 Lefkosia, Cyprus}
\author{E.M. Forgan}
\affiliation{School of Physics and Astronomy, University of
Birmingham, Birmingham B15 2TT, UK}

\author{N. Momono}
\author{M. Oda}
\affiliation{Department of Physics, Hokkaido University, Sapporo 060-0810,
Japan}

\author{C. Dewhurst}
\affiliation{Institut Laue-Langevin, 6 rue Jules Horowitz, B.P. 156-38042,
Grenoble Cedex, France}

\author{C. Baines}
\affiliation{Laboratory for Muon Spin Spectroscopy, PSI Villigen,
CH-5232 Villigen PSI, Switzerland}

\begin{abstract}
The vortex glass state formed by magnetic flux lines in a type-II
superconductor is shown to possess non-trivial three-body
correlations. While such correlations are usually difficult to
measure in glassy systems, the magnetic fields associated with  the
flux vortices allow us to probe these {\em via} muon-spin rotation
measurements of the local field distribution.  
We show {\it via} numerical simulations and analytic
calculations that these observations provide
detailed microscopic insight into the local order
of the vortex glass and more generally validate a
theoretical framework for correlations in glassy systems.
\end{abstract}

\maketitle

In systems which possess long range order, such as
atomic crystals, the local arrangement of particles
is easily obtained from scattering experiments. In
disordered systems, the average correlation between
the positions of two particles can be measured by
scattering techniques, but inferring anything more
about the local geometry is a far more subtle issue.
Little is known experimentally about correlation
functions of higher order.  Measurements of
three-body correlation functions for colloids
imaged in a quasi-two-dimensional geometry have been
reported recently~\cite{maret}.  However, {\em bulk}
measurements of three-body correlation functions in any
system are still unavailable and our understanding of
such correlations derives mainly from simulations. This
Letter reports a study of the local structure of
the vortex glass phase
in a bulk type-II superconductor. The vortex glass phase
provides an example of a glassy system where the local
geometry in the {\em bulk} is uniquely amenable to
investigation, due to the magnetic fields associated
with the vortices, which we measure by the muon-spin
rotation ($\mu$SR) technique. By coupling these
measurements with  Monte Carlo simulations and analytic
calculations, we demonstrate both the existence of 
non-trivial three-body correlations in the flux-line
array and a theoretical framework in which they
may be understood.

In the mixed state of a type-II superconductor,
an applied magnetic field penetrates as lines of
magnetic flux, quantized in units of the flux quantum
$\Phi_0 = h/2e$. Such vortex lines would form an
Abrikosov flux lattice at low temperatures in the
absence of quenched disorder. As the temperature
or the strength of disorder is increased, ordered
arrangements of vortex lines yield to disordered
ones~\cite{review,shobo}. Weak quenched disorder
converts the crystal into a ``Bragg glass"
with quasi-long range order in translational
correlations~\cite{giamarchi}. At stronger disorder,
``vortex glass" states with short-ranged correlations
are obtained.  Neutron scattering measurements
support the proposal of a power-law decay of
translational correlations in the Bragg glass
phase~\cite{bragg_expt}. In contrast, structure
and correlations in vortex glasses remain little
understood.

\begin{figure}
\includegraphics[scale=0.28]{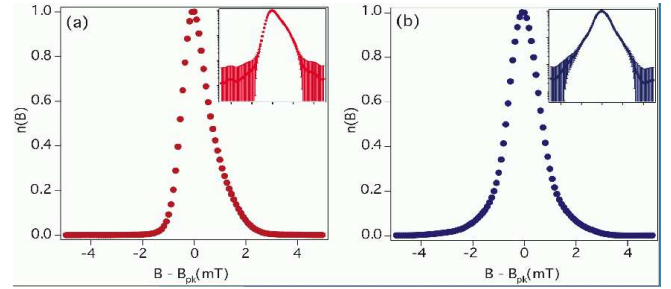}
\caption{(color online) The field distribution function $n(B)$ for
two values of the applied magnetic field H at T=5 K: (a) H = 40 mT
and (b) H = 0.5 T. The insets show the same data, plotted over the
same range, with the y-axis plotted on a logarithmic scale,
indicating the errors on the points. The curves are normalized at
the mode of the distribution, B$_{pk}$. }
\end{figure}

Our experimental system is La$_{1.9}$Sr$_{.1}$CuO$_{4-\delta}$
(LSCO), an underdoped high-T$_c$ superconductor with properties
which amplify the effects of thermal fluctuations and quenched
disorder. It was recently shown using $\mu$SR measurements, on the
same high quality crystal as used in this experiment, that there is
a field induced transition in LSCO to a vortex glass
phase~\cite{divakar}. Our present $\mu$SR experiments were performed
using the GPS spectrometer at PSI, Switzerland with the field nearly
parallel to the $c$-axis of the crystal. The experimental
arrangement was as described in ref.~\cite{divakar}. 
The novelty of the present work is that by relating
the third moment of the magnetic field distribution to an integral
over a three-particle structure factor, we are able to provide
information about three-body correlations in both the ordered and
the glassy phases.

In a $\mu$SR experiment, the probability distribution $n(B)$ of the
spatially varying magnetic field is inferred from the muon
precession signal~\cite{sonier}. This distribution reflects the
arrangement of vortex lines. We measure $n(B)$ as a function of
external magnetic field and temperature, calculating the second
moment $\langle[\Delta B]^2\rangle$, the third moment
$\langle[\Delta B]^3\rangle$, and the related dimensionless
lineshape anisotropy ratio $\alpha = \langle [\Delta
B]^3\rangle^{1/3}/ \langle [\Delta B]^2\rangle^{1/2}$, with the
$k^{th}$ moment defined by $\langle [\Delta B]^k\rangle= \sum
n(B_i)(B_i - \langle B\rangle)^k / \sum n(B_i)$. Figs. 1(a) and (b)
show field distributions for two values of the applied magnetic
field. As discussed in Ref.~\cite{divakar}, the data at 40 mT (Fig.
1(a)) show a tail on the high-field side and hence positive sign of
$\alpha$ expected for a lattice (or Bragg glass) structure. However,
the 0.5 T data (Fig. 1(b)) show a broader and more symmetric
distribution in which this tail is absent and the lineshape at this
field is slightly skewed the opposite way, signalling that the
flux-line structure is not ordered like a lattice. Instead, it is in
a vortex glass state, which dominates the phase diagram in the inset
to Fig. 3.

\begin{figure}
\includegraphics[scale=0.30]{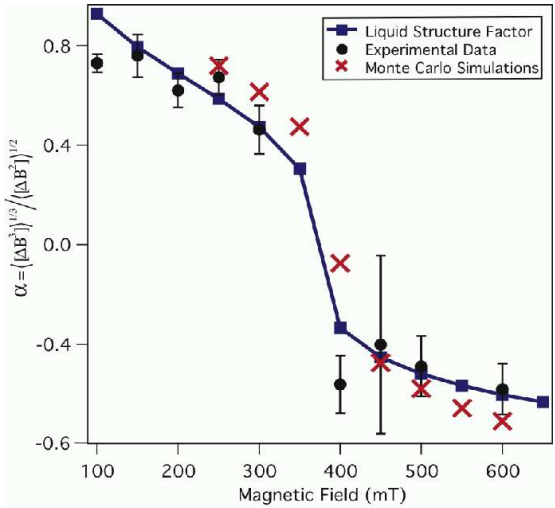}
\caption{(color online) The lineshape anisotropy ratio $\alpha$  as
obtained in (i) experiments at 5K (black circles), (ii) calculated
from Monte Carlo simulations of disordered structures (red crosses)
and from (iii) a theoretical description in terms of liquid state
theory (blue squares connected by a solid line). }
\end{figure}

Fig. 2 shows the behavior of the lineshape anisotropy ratio $\alpha$
as a function of magnetic field at 5 K after field cooling; all
points lie within the vortex glass phase ~\cite{divakar} (inset
Fig.~3).  The experimental data are the sequence of black circles;
these data are modeled theoretically (see below) by the sequence of
red crosses and blue connected squares. Note the reduction in
$\alpha$ beginning at relatively low field values, the precipitous
change of sign at $B \sim 0.35$ T, followed by saturation at an
approximately constant negative value.  Fig.~3 shows the variation
of $\alpha$ over the $H-T$ plane, further illustrating a change of
sign from positive to negative values that occurs {\em deep within}
the vortex glass phase.  We note that a negative third moment of the
field distribution is also observed in the vortex liquid regime;
this is an {\em outstanding problem} for theories of vortex line
correlations~\cite{harshman,blasius}.

\begin{figure}
\hspace{0.5in}
\includegraphics[scale=0.34]{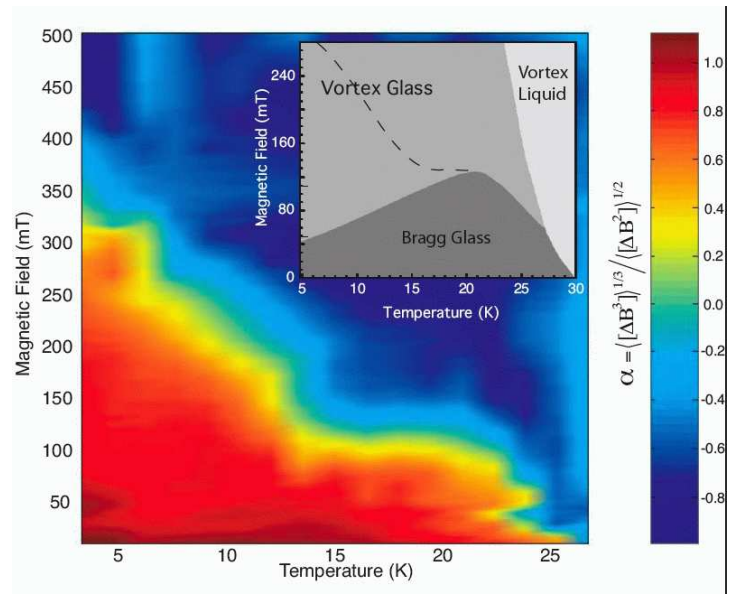}
\caption {(color online) Regimes in H-T space associated with a
fixed sign of $\alpha = \langle [\Delta B]^3\rangle^{1/3}/ \langle
[\Delta B]^2\rangle^{1/2}$, and thus of the third moment. The inset
shows the schematic phase diagram for this system in H-T space,
obtained from a combination of muon-spin rotation, magnetic and Small Angle Neutron Scattering
measurements~\cite{divakar}. The change of sign of $\alpha$ reported
here occurs deep within the vortex glass phase shown in the inset
(dotted line in inset) and over a fairly short range in field and
temperature. }
\end{figure}

For a system of
flux-lines taken to be rigid along the z-axis (field direction)
the third moment is obtained from
\begin{equation}
\langle[\Delta B]^3\rangle \propto \int\!\!\!\int d{\bf q}_1 d{\bf
q}_2 S^{(3)}({\bf q}_1,{\bf q}_2) b({\bf q}_1)b({\bf q}_2)b(-{\bf
q}_1-{\bf q}_2), \label{third}
\end{equation}
where $S^{(3)}({\bf q}_1,{\bf q}_2) = \frac{1}{N} \langle \delta
\rho({\bf q}_1) \delta \rho({\bf q}_2) \delta \rho(-{\bf q}_1-{\bf
q}_2) \rangle$~\cite{hanmac} is the triplet structure factor
and the proportionality constant is
$B/(2\pi)^4\Phi_0$. Here, $b({\bf q})$ is the field of a single
vortex in Fourier space, while $\delta \rho({\bf q})$ is the Fourier
component at wave-vector ${\bf q}$ of the deviation of the flux-line
density from its average value. The second moment is related
similarly to the two-particle correlation function, {\it i.e.} to
the conventional structure factor $S(q) = \frac{1}{N}\langle \delta
\rho({\bf q}) \delta \rho({\bf -q}) \rangle$.  In these expressions,
N is the total number of vortices and the brackets
$\langle\cdots\rangle$ denote an average over the
sample~\cite{brandt3,theory1}. Note that $S(q) \geq 0$.

The simple London model gives $b({\bf q}) = B/(1 + \lambda^2
q_\perp^2)$, where $\lambda$ is the $ab-$plane penetration depth and
${\bf q} = ({\bf q}_\perp, q_z=0)$. However, this yields a magnetic
induction which diverges at the vortex core. This unphysical
divergence is eliminated by multiplying by a ``form factor"
$f(q)$~\cite{yaouanc}. All the available analytic expressions for
$f(q)$ give positive values for all $q$~\cite{cutoff}. For a perfect
triangular lattice, Eq. 1 reduces to a sum over two sets of
reciprocal lattice vectors. Each term in such a sum is manifestly
positive. The form factor reduces the {\em value} of the third
moment, but it cannot change its {\em sign}.

We now relax the requirement of a lattice structure.  One
possibility is to assume uncorrelated lines, i.e. $S(q) = 1$,
$S^{(3)}({\bf q}_1,{\bf q}_2) = 1$~\cite{brandt3}. This limit yields
anomalously large field (B$_0$) dependent values for the second
moment ($\langle[\Delta B]^2\rangle = B_0\Phi_0/4\pi\lambda^2$, in
comparison to the perfect lattice, for which $\langle[\Delta
B]^2\rangle  = 3.71\times 10^{-3} \Phi_0^2/\lambda^4$). It also
leads to a positive third moment. Alternatively, if one applies the
``convolution approximation" from the fluid
literature~\cite{hanmac}, one can express $S^{(3)}$ in terms of the
two-particle correlator only: $S^{(3)}({\bf q_1,q_2}) =
S(q_1)S(q_2)S(|{\bf q_1- q_2}|)$. However, this approach also gives
positive $\langle [\Delta B]^3\rangle$ because the integrand in Eqn.
1 is manifestly positive. The effects of line wandering can be
incorporated using expressions for $S(q_\perp,q_z)$ obtained via the
boson approximation~\cite{review} - numerical values of the moments
are reduced but again the sign of $\langle [\Delta B]^3 \rangle$
should remain positive~\cite{theory1}.

\begin{figure}
\includegraphics[scale=0.35]{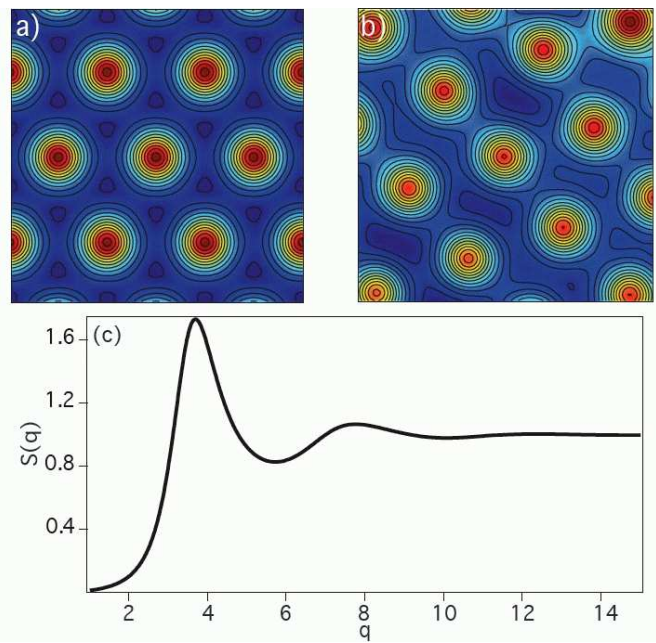}
\caption {(color online) Contours of constant magnetic field
strength B(r) computed for (a) a perfect Abrikosov lattice at B=0.5
T and for (b) a disordered state with a translational correlation
length of about 4 inter-vortex spacings generated using simulated
annealing techniques, with $\gamma/\gamma^\prime = 3$ (see text). In
(c) is shown a plot of the structure factor $S(q)$, obtained
analytically as described in the text ($q$ is given in units of the
inverse mean inter-particle spacing).}
\end{figure}

Negative third moments in vortex glass phases which lack long range
order can thus only result from vortex line arrangements with { \em
non-trivial three-particle correlations} beyond the convolution
approximation. To validate these ideas, we have generated
simulations of disordered states with these attributes. Simulated
annealing techniques were used to find the ground state of a system
of 6400 particles (rigid vortex lines) interacting through a
potential of the form $\gamma K_0(r/\lambda)$, and with a quenched
one-body potential arising from 3700 randomly placed pinning sites.
The pinning sites were modeled as attractive Gaussian wells of the
form $V_d(r) = -\gamma^\prime e^{-r^2/R_p^2}$, where $R_p$ is a
length scale for disorder and $\gamma^\prime$ gives the strength of
the disorder. We took $R_p = 0.1$ in units of the inter-particle
separation $a$, and varied $\gamma/\gamma^\prime$ in the range
[0.001:1000]. Without disorder, the ground state is a crystal. As
the strength of pinning is increased, this crystal fragments into
smaller domains~\cite{theory2} in a polycrystalline arrangement in
which the typical domain size is controlled by the pinning strength
~\cite{chandran}.  We can then calculate $n(B)$ and its moments by
constructing a histogram of local field values~\cite{gridding}. In
Fig. 4, we compare field contours obtained for a perfect triangular
lattice (a), and a disordered arrangement with the attributes
discussed above (b). The disordered case shows the absence of long
range order in both vortex position and field value, while
maintaining a marked tendency towards local triangular coordination.
Explicit calculations of $\alpha$ from a distribution of vortex
lines with such a structure are a very good representation of the
experimental data in Fig. 2.

The same general result may be obtained by analytic calculations in
certain limits. The triplet structure factor is conventionally
decomposed as
\begin{equation}
S^3({\bf q}_1,{\bf q}_2)=
S(q_1)S(q_2)S(|-{\bf q}_1-{\bf q}_2|)[1 +
\rho^2C^{(3)}({\bf q}_1,{\bf q}_2)],
\end{equation}
where $C^{(3)}$ is the triplet direct correlation
function~\cite{hanmac} and $\rho$ is the average density of the
liquid~\cite{theory1}. In our earlier decomposition of $S^{(3)}({\bf
q_1,q_2})$, using the convolution approximation, we assumed
$C^{(3)}=0$, an approximation which is clearly inadequate
here~\cite{convolution}. To progress beyond the convolution
approximation, we assume that the lines are rigid over the scale of
the penetration depth and we model translational correlations deep
in the vortex glass phase in terms of correlations in an equilibrium
fluid~\cite{theory1}, using an accurate analytic approximation for
$C^{(3)}({\bf q}_1,{\bf q}_2)$~\cite{zhou}. We use the pair
correlations of the $\gamma K_0(r/\lambda)$ potential appropriate to
rigid flux lines to describe local structure in the glass phase.
These are obtained from solutions to the self-consistent equations
of classical liquid state theory in the hyper-netted chain
approximation~\cite{hanmac}.  Our model $S(q)$ (Fig. 4(c)) is
derived from liquid state computations at a single density
($\lambda/a \sim 4.0$, with $a$ the mean inter-particle spacing,
corresponding to a field of 0.5 T and $\lambda \sim 2800$ \AA) using
a value for the coupling constant $\Gamma = \gamma/k_B T$ of 50. 
We assume that the local structure of the
glass as captured in $S(q)$ is not altered substantially as the
field is varied, once all length scales are expressed in units of
the mean inter-particle spacing $a$, an assumption which should be
valid in the limit where $\lambda \gg a$.  Using Eqs. 1 and 2, we
calculate $\alpha$, illustrated in Fig. 2 as the sequence of
connected blue squares. Note that as the field value is increased,
the third moment changes sign, with $\alpha$ saturating at a value
of about -0.6, close to the value in the experimental data. This
relatively simple analytical model thus enables a robust description
of the effects of three-body correlations in the vortex glass phase.

Similar behavior is also seen in the highly anisotropic  superconductor
Bi$_2$Sr$_2$CaCu$_2$O$_{8+\delta}$ over a range of doping~\cite{harshman,blasius}, suggesting that  system-specific
interpretations of the negative third moment are  unlikely. In ref.~\cite{gilardi, drew}
a transition from a triangular to a square  vortex lattice was observed
in the more highly doped system. We find  no evidence for this at the
field values we probe. At high fields our  $\mu$SR data do not show the
tail in n(B) on the high field side  which would signal a crystalline
arrangement, whether square or  triangular. Magnetic order can coexist
with superconductivity in LSCO  over a restricted doping range. We have
characterized the sample with  neutrons and with longitudinal $\mu$SR,
finding magnetic signatures  only below about 4 K~\cite{epaps}. We therefore restrict
ourselves here to  temperatures above this value where any magnetic
fluctuations, should  they exist, lie well outside the muon time window,
and cannot  contribute to the depolarization.

The sign of the third moment reflects the competition
between the (positive) contributions from the vortex
cores, which yield the positive tail of $n(B)$,
and (negative) contributions from field values at
the centres of the triangles formed locally by the
vortices and associated with the minima of $n(B)$.
Structures with strong local triplet correlations but
no long-range order protect both these contributions,
but subtly enhance the negative ones, since now the
positions of particles can fluctuate relative to each
other, unlike in the perfect crystal, while retaining a
strong tendency to local triangular coordination (see
Fig. 4(b)) as in the crystal. The sign changes arises 
when the negative contributions to the integral 
(Eq. 1) overwhelm the positive contributions.  
The integrand of Eq. 1 varies strongly as a
function of $q$ in the { \em vicinity of the first
peak of $S(q)$ and below}. For larger $q$, form factor
cut-offs set in, while at smaller $q$, the integrand
becomes negative, due to the generally large and
{\em negative} value of $C^{(3)}$ in this region
(see Ref.~\cite{zhou}). The resultant sign depends
on the location of the first peak of $S(q)$, which
is itself determined by the magnetic field. Thus,
non-trivial three-body correlations arise out of
$C^{(3)}$ in a disordered system (which, unlike a
crystal, has contributions to the integral at smaller
$q$ than the first Bragg peak i.e. at length scales
which are a little larger than nearest neighbor
vortex spacings). Although the change in sign of the
third moment at high fields confirms the existence
of non-trivial three-body correlations, it does not
indicate the formation of a new vortex state. Instead,
this observation supports 
our simple theoretical and computational models of
structure and correlations in the vortex glass state,
from which such a change in sign follows naturally as
the magnetic field is varied, without the requirement
of a phase transition.

In conclusion, this paper describes an unusual
experimental consequence of many-particle correlations
in a magnetic flux line system, showing how
three-body correlations are responsible for negative
third moments in the field distributions associated with
glassy phases of vortex lines. 
Our results motivate and validate the use of
simple analytic approximations to describe three-body
correlations in bulk disordered systems, an approach
which should find wider application in areas outside
the field of superconductivity.

This work was supported by the DST(India), the Swiss
National Foundation and NCCR MaNEP, the EPSRC(UK) and
the Ministry of Education, Science and Technology of
Japan. The $\mu$SR experiments were performed at the
Swiss Muon Source, Paul Scherrer Institute, Villigen.

\vspace{5mm}


\begin{thebibliography}{99}
\bibitem{maret} K. Zahn {\it et al.}, Phys. Rev. Lett. {\bf 91} 115502 (2003).
\bibitem{review} G. Blatter {\it et al.}, Rev. Mod. Phys, {\bf 66}, 1125 (1994).
\bibitem{shobo} T. Giamarchi and S. Bhattacharya, in
{\em High Magnetic Fields: Applications in Condensed Matter Physics and
Spectroscopy}, ed. C. Berthier {\it et al.}, Springer-Verlag, (2002), p 314.
\bibitem{giamarchi} T. Giamarchi and P. Le Doussal, Phys. Rev. Lett.
{\bf 72}, 1530 (1994).
\bibitem{bragg_expt} T. Klein {\it et al.}, Nature {\bf 413}, 404 (2001).
\bibitem{divakar} U. Divakar {\it et al.},  Phys. Rev. Lett. {\bf 92},
237004, (2004).
\bibitem{sonier} J.E. Sonier, J. H. Brewer and R.F. Kiefl, Rev.
Mod. Phys. {\bf 72}, 769 (2000).
\bibitem{harshman} D.R. Harshman {\it et al.}, Phys. Rev. Lett,
{\bf 67} 3152 (1991).
\bibitem{blasius} T. Blasius {\it et al.}, Phys. Rev. Lett., {\bf 82} 4926 (1999).
\bibitem{hanmac} J. P. Hansen and I. R. Macdonald, {\em
Theory of Simple Liquids}, (Academic, London, 1986), 2nd edition.
\bibitem{brandt3} E. H. Brandt, Phys. Rev. Lett., {\bf 66}, 3213 (1991)
\bibitem{theory1} G. I. Menon {\it et al.},  Phys. Rev. B, {\bf 60}, 7607, (1999)
\bibitem{yaouanc} A. Yaouanc {\it et al.}, Phys. Rev. B, {\bf 55}, 11107 (1997).

\bibitem{cutoff} Analytic expressions for $f(q)$ include $f(q) = g K_1(g)$, where $g
= \sqrt{2}\xi(q^2 + \lambda^{-2})^{1/2}$ and $K_1(x)$ is a modified Bessel function, the Gaussian form
$f(q) = \exp(-\xi^2q^2/2)$ and the exponential $f(q) = \exp(-\sqrt{2}\xi q)$~\cite{yaouanc}.

\bibitem{theory2} G. I. Menon, Phys. Rev. B, {\bf 65},
104527 (2002)
\bibitem{chandran}M. Chandran~{\it et al.}, Phys. Rev. B {\bf 69}, 024526 (2004).
\bibitem{gridding} We evaluate the local field
at $10^6$ points chosen at random within the
simulation box, calculating $n(B)$ and its moments
from these measurements.
\bibitem{convolution} H. W. Jackson and E. Feenberg,  Rev. Mod.
Phys, {\bf 34}, 683 (1962)
\bibitem{zhou} S. Zhou and E. Ruckenstein, Phys. Rev. E,
{\bf 61}, 2704 (2000).

\bibitem{gilardi} R. Gilardi~{\it et al.}, Phys. Rev. Lett {\bf 88}, 217003 (2002) 
\bibitem{drew} A. J. Drew~{\it et al.}, Physica B {\bf 374}, 203 (2006)
\bibitem{epaps}See EPAPS Document No. [E-PRLTAO-97-105644] for a
supplemental figure. For more information on EPAPS, 
see http://www.aip.org/pubservs/epaps.html.

\end{thebibliography}
\end{document}